\documentclass[aps,twocolumn,prb,eqsecnum,showpacs]{revtex4}
\usepackage[dvipdfm]{graphicx}
\begin{document}
\draft
\preprint{\today}
\title{Photoinduced Infrared Absorption of Quasi-One-Dimensional
       Halogen-Bridged Binuclear Transition-Metal Complexes}
\author{Jun Ohara and Shoji Yamamoto}
\address{Division of Physics, Hokkaido University,
         Sapporo 060-0810, Japan}
\date{\today}
\begin{abstract}
We investigate the optical conductivity of photogenerated solitons in
quasi-one-dimensional halogen-bridged binuclear transition-metal
($M\!M\!X$) complexes with particular emphasis on a comparison among the
three distinct groups:
 $A_4$[Pt$_2$(P$_2$O$_5$H$_2$)$_4X$]$\cdot$$n$H$_2$O
($X=\mbox{Cl},\mbox{Br},\mbox{I}$;
 $A=\mbox{Na},\mbox{K},\mbox{NH}_4,\cdots$),
 Pt$_2$($R$CS$_2$)$_4$I
($R=\mbox{C}_n\mbox{H}_{2n+1}$) and
 Ni$_2$(CH$_3$CS$_2$)$_4$I,
which exhibit a mixed-valent ground state with the $X$ sublattice
dimerized, that with the $M_2$ sublattice dimerized and a Mott-Hubbard
magnetic ground state, respectively.
Soliton-induced absorption spectra for
$A_4$[Pt$_2$(P$_2$O$_5$H$_2$)$_4X$]$\cdot$$n$H$_2$O should split into two
bands, while that for Pt$_2$($R$CS$_2$)$_4$I and
Ni$_2$(CH$_3$CS$_2$)$_4$I should consist of a single band.
The excitonic effect is significant in Ni$_2$(CH$_3$CS$_2$)$_4$I.
\end{abstract}
\maketitle

   $M\!X$ chains, that is, a family of quasi-one-dimensional halogen
($X$)-bridged transition-metal ($M$) complexes, provide an exciting stage
\cite{N3865,G6408,W6435,Y422} performed by electron-electron correlation,
electron-lattice interaction, low dimensionality and $d$-$p$ orbital
hybridization.
In recent years, binuclear metal analogs which are referred to as
$M\!M\!X$ chains have stimulated renewed interest
\cite{K533,Y125124} in this system.
The existent $M\!M\!X$ compounds consist of three groups:
 $A_4$[Pt$_2$(pop)$_4X$]$\cdot$$n$H$_2$O
($X=\mbox{Cl},\mbox{Br},\mbox{I}$;
 $A=\mbox{Li},\mbox{Cs},\cdots$;
 $\mbox{pop}=\mbox{diphosphonate}
 =\mbox{P}_2\mbox{O}_5\mbox{H}_2^{\,2-}$) \cite{C4604,C409},
 Pt$_2$($R$CS$_2$)$_4$I
($R=\mbox{alkyl\ chain}
 =\mbox{C}_n\mbox{H}_{2n+1}$) \cite{B444,I115110}
and
 Ni$_2$(dta)$_4$I
($\mbox{dta}=\mbox{dithioacetate}
 =\mbox{CH}_3\mbox{CS}_2^{\,-}$) \cite{B2815}.
$A_4$[Pt$_2$(pop)$_4X$]$\cdot$$n$H$_2$O resembles conventional $M\!X$
compounds and generally exhibits a similar ground state of mixed valence
with halogen-sublattice dimerization \cite{K40,B1155},
which is referred to as the charge-density-wave (CDW) state.
Pt$_2$($R$CS$_2$)$_4$I exhibits a distinct ground state with
metal-sublattice dimerization \cite{K10068},
which is referred to as the alternate charge-polarization (ACP) state.
Ni$_2$(dta)$_4$I exhibits the averaged-valence state without any lattice
distortion \cite{S265} and is regarded as an insulator of the
Mott-Hubbard (MH) type.
These ground states can be tuned by pressure
\cite{S66,Y140102,M046401,I2149} as well as replacing the bridging
halogens \cite{B1155,Y1198}, counter ions \cite{M046401,Y2321} and
ligand molecules \cite{I387}.

   In such circumstances, topological excitations \cite{Y189} such as
solitons and polarons have been found for an $M\!M\!X$ Hamiltonian of the
Su-Schrieffer-Heeger type \cite{S1698} and an analogy between $M\!M\!X$
chains and {\it trans}-polyacetylene has been pointed out.
The direct $M(d_{z^2})$-$M(d_{z^2})$ overlap effectively reduces the
on-site Coulomb repulsion and therefore electrons can be more itinerant in
$M\!M\!X$ chains.
In fact $M\!M\!X$ chains exhibit much higher room-temperature conductivity
than $M\!X$ chains \cite{K10068}.
Solitons generally have lower formation energies and smaller effective
masses than polarons in $M\!M\!X$ chains \cite{Y165113}.
Then we take more and more interest in $M\!M\!X$ solitons as charge or
spin carriers.
While $M\!M\!X$ uniform absorption spectra have recently been investigated
\cite{K2163}, but photoinduced ones, which serve as prominent probes for
nonlinear excitations, have neither measured nor calculated yet.
Thus motivated, we study optical conductivity for $M\!M\!X$ solitons and
illuminate a possible contrast among the family compounds.

   We describe $M\!M\!X$ chains by the one-dimensional
$\frac{3}{4}$-filled single-band Peierls-Hubbard Hamiltonian
\begin{eqnarray}
   &&
   {\cal H}
   =-\sum_{n,s}
     \Bigl\{
      t_{M\!M}b_{n,s}^\dagger a_{n,s}
     +\bigl[t_{M\!X\!M}
   \nonumber \\
   &&\qquad
      -\alpha(v_{n+1}-v_n)\bigr]a_{n+1,s}^\dagger b_{n,s}
     +\mbox{H.c.}
     \Bigr\}
   \nonumber \\
   &&\qquad
    -\beta\sum_{n,s}
     \bigl[
      (v_n-u_{n-1})n_{n,s}+(u_n-v_n    )m_{n,s}
     \bigr]
   \nonumber \\
   &&\qquad
    +U_{M}\sum_{n}(n_{n,+}n_{n,-}+m_{n,+}m_{n,-})
   \nonumber \\
   &&\qquad
    +\sum_{n,s,s'}
     (V_{M\!M}n_{n,s}m_{n,s'}+V_{M\!X\!M}n_{n+1,s}m_{n,s'})
   \nonumber \\
   &&\quad
   +\frac{K_{M\!X}}{2}\sum_{n}
    \bigl[(u_n-v_n)^2+(v_n-u_{n-1})^2\bigr],\hspace{1cm} (1) \nonumber
\end{eqnarray}
\begin{figure}
\centering
\includegraphics[width=74mm]{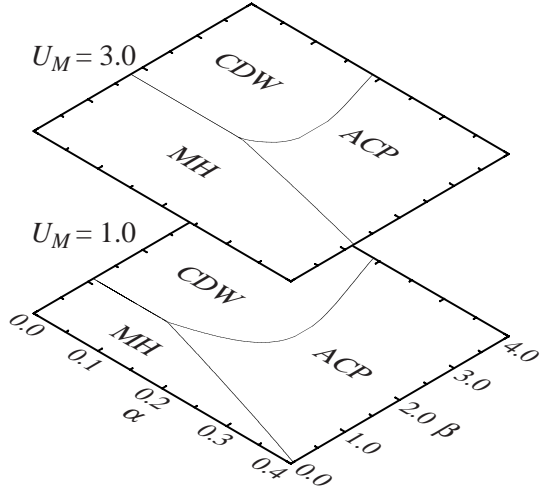}
\vspace*{1mm}
\caption{Ground-state phase diagrams.}
\label{F:PhD}
\end{figure}
\noindent
where
$n_{n,s}=a_{n,s}^\dagger a_{n,s}$ and
$m_{n,s}=b_{n,s}^\dagger b_{n,s}$ with
$a_{n,s}^\dagger$ and $b_{n,s}^\dagger$ being the creation
operators of an electron with spin $s=\pm$ (up and down) for the
$M\,d_{z^2}$ orbitals in the $n$th $M\!M\!X$ unit.
$t_{M\!M}$ and $t_{M\!X\!M}$ describe the intradimer and interdimer
electron hoppings, respectively.
$\alpha$ and $\beta$ are the intersite and intrasite electron-lattice
coupling constants, respectively, with $K_{M\!X}$ being the
metal-halogen spring constant.
$u_n$ and $v_n$ are, respectively, the chain-direction displacements
of the halogen and metal dimer in the $n$th $M\!M\!X$ unit from their
equilibrium positions.
We assume, based on the thus-far reported experimental observations,
that every $M_2$ moiety is not deformed.
The lattice distortion is adiabatically determined through the
Hellmann-Feynman force equilibrium condition.
We show in Fig. \ref{F:PhD} the ground-state phase diagrams within the
Hartree-Fock (HF) approximation.
Setting $t_{M\!X\!M}$ and $K_{M\!X}$ both equal to unity, we adopt a
common parameter set: $t_{M\!M}=2.0\,,V_{M\!M}=0.5\,,V_{M\!X\!M}=0.3$.
The rest are taken in three ways as
a) $U_{M}=1.0\,,\alpha=0.0\,,\beta=1.4$;
b) $U_{M}=1.0\,,\alpha=0.3\,,\beta=1.0$;
c) $U_{M}=3.0\,,\alpha=0.3\,,\beta=1.0$;
which are relevant to
$A_4$[Pt$_2$(pop)$_4X$]$\cdot$$n$H$_2$O, Pt$_2$($R$CS$_2$)$_4$I and
Ni$_2$(dta)$_4$I, and indeed give the CDW, ACP and MH ground states,
respectively.

   The optical spectra are obtained by calculating the matrix elements
between the ground state $|{\rm g}\rangle$ of energy $E_{\rm g}$ and the
excited states $|l\rangle$ of energy $E_l$ ($l=1,2,\cdots$) for the
current operator
\begin{eqnarray}
   &&
   {\cal J}
   =\frac{{\rm i}e}{\hbar}\sum_{n=1}^N\sum_{s=\pm}
    \Bigl\{
     c_{M\!M}t_{M\!M}
     (b_{n,s}^\dagger a_{n,s}-a_{n,s}^\dagger b_{n,s})
   \nonumber\\
   &&\qquad
    +c_{M\!X\!M}\bigl[t_{M\!X\!M}-\alpha(v_{n+1}-v_n)\bigr]
   \nonumber\\
   &&\qquad\times
     (a_{n+1,s}^\dagger b_{n,s}-b_{n,s}^\dagger a_{n+1,s})
    \Bigr\},\hspace{2.5cm} (2) \nonumber
\end{eqnarray}
where $c_{M\!M}$ and $c_{M\!X\!M}$ are the average $M$-$M$ and $M$-$X$-$M$
distances, respectively, and are set for $c_{M\!X\!M}=2c_{M\!M}$.
The real part of the optical conductivity is given by
\begin{eqnarray}
   &&
   \sigma(\omega)
    =\frac{\pi}{N\omega}\sum_l
   |\langle l|{\cal J}|{\rm g}\rangle|^2
   \delta(E_l-E_{\rm g}-\hbar\omega).
\hspace{1.1cm} (3) \nonumber
\end{eqnarray}
$|{\rm g}\rangle$ is set for the HF ground state, while $|l\rangle$ is
calculated within and beyond the HF approximation, being generally defined
as
\begin{eqnarray}
   &&
   |l\rangle
   =\sum_s
    \sum_{\epsilon_\mu\leq\epsilon_{\rm F}}
    \sum_{\epsilon_\nu>\epsilon_{\rm F}}
    f(\mu,\nu,s;l)c_{\nu,s}^\dagger c_{\mu,s}|{\rm g}\rangle,
\hspace{1.3cm} (4) \nonumber
\end{eqnarray}
where $\epsilon_{\rm F}$ is the Fermi energy and $c_{\lambda,s}^\dagger$
creates an electron with spin $s$ for the $\lambda$th HF eigenstate with
an eigenvalue $\epsilon_\lambda$.
In the HF scheme, any excited state is simply approximated by a single
Slater determinant as $f(\mu,\nu,s;l)=\delta_{\mu\nu s,l}$.
We further consider excited states of the configuration-interaction (CI)
type, where $f(\mu,\nu,s;l)$ is determined so as to diagonalize the
original Hamiltonian (1).
We set $N$ equal to $84$, where $2$ GB memory is necessary for the CI
calculation.

\begin{figure}
\centering
\includegraphics[width=72mm]{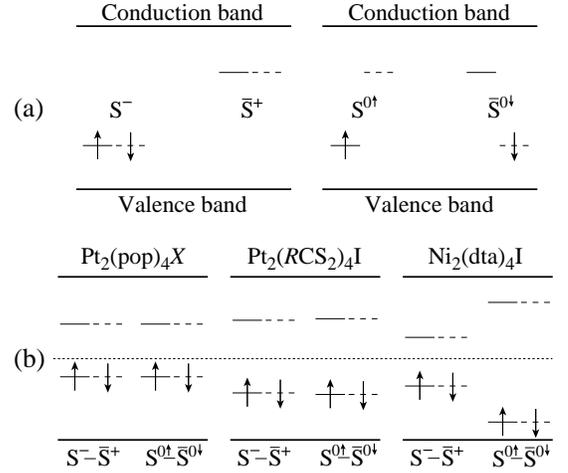}
\vspace*{1mm}
\caption{Localized levels due to solitons in a pair, where the solid and
         broken segments designate those for up-spin ($\uparrow$) and
         down-spin ($\downarrow$) electrons, respectively.
         (a) Qualitative illustrations for charged and neutral soliton
             pairs.
         (b) Quantitative illustrations for soliton pairs in
             $A_4$[Pt$_2$(pop)$_4X$]$\cdot$$n$H$_2$O,
             Pt$_2$($R$CS$_2$)$_4$I and Ni$_2$(dta)$_4$I,
             where the dotted line designates the center of the gap.}
\label{F:level}
\end{figure}

   Since photogenerated defects are necessarily in pairs, we find the most
stable soliton (S)-antisoliton ($\bar{\mbox{S}}$) pair at a sufficiently
low temperature $k_{\rm B}T/t_{M\!X\!M}=10^{-3}$ without any assumption on
the soliton shapes.
A pair of solitons generally gives two additional levels within the gap,
as is illustrated in Fig. \ref{F:level}(a).
There appear further soliton-related intragap levels in the
strong-coupling region \cite{T4074}.
These levels are strongly localized and completely assigned to each
soliton.
The spin up-down symmetry holds with a charged soliton pair
$\mbox{S}^-$$-$$\bar{\mbox{S}}^+$, whereas it breaks down with a neutral
soliton pair $\mbox{S}^{0\uparrow}$$-$$\bar{\mbox{S}}^{0\downarrow}$.
The lower level is doubly filled, while the upper one is vacant.
Considering that there is no essential overlap between the wave functions
of well-separated $\mbox{S}$ and $\bar{\mbox{S}}$, either an optical
transition between the vacant soliton level and the valence band or that
between the filled soliton level and the conduction band may yield
absorption in the gap.
The level structure of any soliton pair quantitatively depends on the
background compound and typical energy schemes are predicted in Fig.
\ref{F:level}(b).
Since the electron-hole symmetry is broken in the Hamiltonian (1),
soliton-related electron and hole levels may be asymmetric with respect to
the center of the gap.
It is indeed the case with the CDW background of
$A_4$[Pt$_2$(pop)$_4X$]$\cdot$$n$H$_2$O, whereas on the ACP background of
Pt$_2$($R$CS$_2$)$_4$I and the MH background of Ni$_2$(dta)$_4$I,
a soliton of charge $\sigma$, S$^{\sigma}$, and that of spin
$s$, S$^{0s}$, described in terms of electrons are still nearly equivalent
to their counterparts S$^{-\sigma}$ and S$^{0\,-s}$ described in terms of
holes, respectively.

\begin{figure*}
\centering
\includegraphics[width=160mm,bb=0 0 766 364]{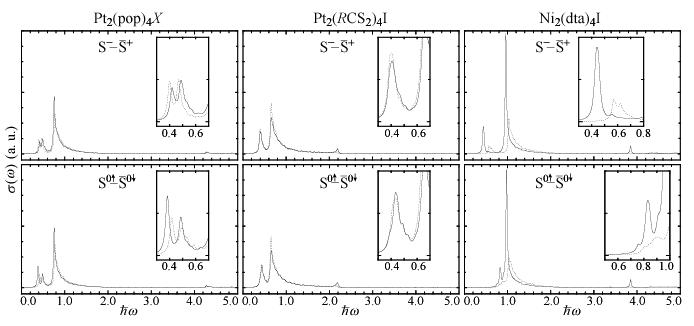}
\vspace*{1mm}
\caption{Hartree-Fock (HF) and single-excitation configuration-interaction
         (SECI) calculations of the optical conductivity spectra for the
         optimum soliton-antisoliton pairs on the CDW, ACP and MH
         backgrounds, where each line has been Lorentzian-broadened.
         The mid-gap absorption bands due to solitons are scaled up in
         insets.}
\label{F:OC}
\end{figure*}

   In consequence, {\it photoinduced soliton absorption spectra for 
$A_4$[Pt$_2$(pop)$_4X$]$\cdot$$n$H$_2$O should split into two bands, while
those for Pt$_2$($R$CS$_2$)$_4$I and Ni$_2$(dta)$_4$I should consist of a
single band}, as is demonstrated in Fig. \ref{F:OC}.
The excitonic effect is remarkable in the strong-correlation compound
Ni$_2$(dta)$_4$I.
In mixed-valent $M\!X$ chains, neutral solitons seem to be the
lowest-energy pair excitations \cite{I1088,O115112}.
In mixed-valent $M\!M\!X$ chains, on the other hand, charged solitons may
be the lowest-energy excitations because the on-site Coulomb repulsion
$U_M$ and the Holstein coupling $\beta$ are effectively smaller and
larger, respectively \cite{Y125124,K10068}.
Polarons have much higher formation energies for both $M\!X$
\cite{G6408,I1088} and $M\!M\!X$ \cite{Y165113} chains and can therefore
be generated from relatively high-energy excited states corresponding to
the electron-hole continuum \cite{M5763}.
An excitation energy close to the Peierls gap directly induces
charge-transfer excitons \cite{M5758} and they may relax into soliton
pairs for $M\!M\!X$ chains as well.
The optical conductivity spectra for
$A_4$[Pt$_2$(pop)$_4X$]$\cdot$$n$H$_2$O clearly distinguish between
charged and neutral solitons.
The charged (neutral)-soliton mid-gap absorption spectrum is double-peaked
and the higher (lower)-energy band has larger oscillator strength.
As for Pt$_2$(dta)$_4$I and Ni$_2$(dta)$_4$I, further measurements such as
electron spin resonance \cite{T2169} are supplementary to find out which
kind of solitons.

   Besides the optimum soliton solutions, there are some metastable ones
in general \cite{Y165113,O}.
In Ni$_2$(dta)$_4$I, their energy difference is relatively large and
therefore the absorption spectra may significantly vary with the soliton
type.
Photoinduced infrared absorption measurements on $M\!M\!X$ compounds
must provide rich physics and are thus encouraged.

   The authors are grateful to K. Iwano, Y. Shimoi and H. Okamoto for
fruitful discussions and helpful comments.
This work was supported by the Ministry of Education, Culture, Sports,
Science and Technology of Japan and the Iketani Science and Technology
Foundation.

\end{document}